*Chapter 7*

# *AB INITIO* STUDY OF ACID MOLECULES INTERACTING WITH $H_2O$

***Aleksey A. Zakharenko***
Center for Superfunctional Materials, Department of Chemistry,
Pohang University of Science and Technology, Pohang, Korea (E-mail:aazaaz@inbox.ru)

## ABSTRACT

Using the Gaussian-03 for *ab initio* calculations, interactions of various acid molecules with a single water molecule were studied. The molecular and supermolecular optimized structures were found with the Becke-3-Lee-Yang-Parr (B3LYP-hybrid potential) calculations of density-functional theory (DFT) methods, as well as the Møller-Plesset second-order perturbation theory using the basis set of aug-cc-*p*VDZ quality and the CRENBL ECP effective core potential for molecules containing heavy iodine atom. Possible isomers of studied acids and supermolecules consisting of acid molecules coupled with a single water molecule are shown. Energies, zero-point energies (ZPEs), thermal enthalpies, and thermal free energies, as well as the corresponding binding energies for the theoretical methods were calculated. It was also found that optimized structures of supermolecular isomers with lowest energies corresponding to the global minimum on the potential energy surfaces can be different for both theories. The most simply structured acids $H_2S$ and $H_2Se$, forming acid-water supermolecules, can give clear evidence of disagreement of the two theoretical methods concerning optimization of lowest energy structures, because the B3LYP-DFT method gives the lowest-energy structure for the first supermolecular isomer, but the MP2 method for the second possible isomer. A dramatic difference was also found between potential energy surfaces for both theories applying to the finding of the optimized structures of the $H_2SO_3$-$H_2O$ supermolecular isomers, because MP2 supermolecular geometries cannot exist for the corresponding B3LYP-DFT ones, for which the frequency characteristics of the supermolecular isomers were also calculated. In general, the binding energies and ZPE ones for the MP2 method are 10 to 15% larger than those for the B3LYP-DFT method. However, the thermal free energies for the MP2 method can be significantly smaller than those for the B3LYP-DFT method. The corrected energies with the basis set superposition error (BSSE) correction were also calculated for the studied acid super-molecules.

## 1. Introduction

One of recurrent topics of physical chemistry and chemical physics is the theoretical investigation of gas-phase clusters [1-4], including hydrogen-bonded complexes. The interaction of an acid molecule with water can lead to many different structures which can depend on chosen level of theory. It is well-known that the hydration process has important implications in the context of atmospheric chemistry and aerosol, because there occur reaction mechanisms of introduction of substantial amount of gas phase chlorine and bromine compounds into the marine troposphere [5-6]. The development of supersonic jet nozzles allowed an extensive study of different molecular clusters in complicated experiments, for instance see Refs. [7-11]. On the other hand, *ab initio* theoretical studies are widely used in addition to experimental investigations. Thus, a large set of the structural and thermochemical data was obtained in this theoretical study with quantum-mechanical methods.

In Ref. [12], it was mentioned that the available structural, spectroscopic, and thermochemical data are still limited for the majority of hydrated halides. That is also true for chlorine-, bromine-, and iodine-containing acids, of which the recent theoretical and experimental investigations can be found in Refs. [13-29]. Hypochlorous (HClO), hypobromous (HBrO), and hypo-iodous (HIO) acids are probably the simplest examples and are weak acids. Chlorous ($HClO_2$), bromous ($HBrO_2$), and iodous ($HIO_2$) acids, pertaining to the relatively weak-acid family, are also well-known. Chloric acid (chemical formula $HClO_3$) is known as a strong acid, $pK_a \sim -1$, and oxidizing agent. Bromic acid with the chemical formula $HBrO_3$ is a key reagent in the well-known Belousov-Zhabotinsky oscillating reaction and has about 62% bromine, 1% hydrogen, and 37% oxygen. Iodic acid with the chemical formula $HIO_3$ can be obtained as a white solid and is an insoluble compound, unlike chloric acid or bromic acid. Perchloric acid ($HClO_4$) is an oxoacid of chlorine and is a colorless liquid soluble in water. Perchloric acid is a strong super-acid completely dissociating in an aqueous solution comparable in strength to sulfuric acid ($H_2SO_4$) or nitric acid ($HNO_3$). Perbromic acid ($HBrO_4$) is a strong acid which strongly oxidizes and is the least stable of the halogen (VII) acids. Per-iodic acid ($HIO_4$) has a heavy iodine atom and is widely used in organic chemistry for structural analysis.

On the other hand, there have been many experimental and theoretical investigations of strong nitric super-acid ($HNO_3$) which is a highly corrosive and toxic acid and can cause severe burns. Interesting example theoretical and experimental studies of nitric acid can be found in Refs. [30-36] and [37-42], respectively. Also, there exists much theoretical (in general, density-functional theory or DFT) and experimental work studying the strong sulfur-containing super-acid, $H_2SO_4$, see Refs. [43-46] and DFT-studies in Refs. [47-50]. It is thought that due to experimental difficulties in investigations of Sulfurous acid $H_2SO_3$, less attention is paid to the acid study [51-52], and little is known about Sulfonic acid, which has the same chemical formula $H_2SO_3$. The sulfur-containing acid with the simplest chemical formula $H_2S$ (Sulfane or Hydrogen sulfide) is a covalent hydride structurally related to water ($H_2O$) because oxygen and sulfur occur in the same group of the periodic table of chemical elements. Hydrogen sulfide is weakly acidic, and references on the acid studies can be found in Refs. [53-56]. Hydrogen selenide ($H_2Se$) is the simplest hydride of selenium studied in Ref. [57], a colorless, flammable gas under standard conditions, and soluble in water. It is noted that the properties of $H_2S$ and $H_2Se$ are similar.

The well-known phosphorous-containing acids, phosphoric acid ($H_3PO_4$) and phosphorous acid ($H_3PO_3$) were studied in Refs. [58-64]. Phosphoric acid ($H_3PO_4$) also known as orthophosphoric acid or phosphoric (V) acid, is an inorganic acid which is very often used as an aqueous solution. Phosphoric acid is also used as the electrolyte in phosphoric-acid fuel cells. Phosphorous acid, also called phosphonic acid, is one of the well-known and commonly-used oxoacids of phosphorus. The other well-known acids are boric acid ($H_3BO_3$ or $B(OH)_3$) also called boracic acid or orthoboric acid, and boron oxide hydroxide (metaboric acid) with the chemical formula $HBO_2$. There is much work on the $H_3BO_3$ and $HBO_2$ acids used in medicine, for instance see Refs. [65-70]. A highly valuable basis for many chemical compounds ranging from polymers to pharmaceuticals, hydrogen cyanide is a chemical compound with the following chemical formula HCN. A solution of hydrogen cyanide in water is called hydrocyanic acid. Hydrogen cyanide is a colorless, very poisonous, and highly volatile liquid which boils slightly above room temperature at 26 °C (78.8 °F). Recent studies on the weakly acidic hydrogen cyanide can be found in Refs. [71-73]. It is noted that HCN as the simplest nitrile system can act as both proton donor and proton acceptor systems in cluster formation that can occur for water and alcohol molecules. It is thought that the interaction of HCN with water can lead to two different structures.

## 2. CALCULATIONS

All the calculations were performed using the GAUSSIAN-03 program [74], see also the famous books [75-77]. The minimum-energy molecular structures of all supermolecular isomers were completely optimized by using density-functional theory (DFT [78]) calculations, employing Becke's three-parameter exchange potential [79-80] and the Lee-Yang-Parr correlation functional [81] (B3LYP) as well as the Møller-Plesset [82] second-order perturbation theory (MP2) calculations, employing Dunning's augmented basis set [83-84] of the aug-cc-*p*VDZ quality. In a super-molecule consisting of two molecules (for instance, in the water dimer) basis functions from one molecule can help to compensate for the basis set incompleteness on the other molecule, and vice versa. This effect is known as the basis set superposition error (BSSE) which will be zero in the case of a complete basis set. An approximate way of assessing the BSSE is the counterpoise (CP) correction method [85-86], in which the BSSE can be estimated as the difference between monomer energies with the regular basis and the energies calculated with the full set of basis functions for the whole super-molecules studied in this paper. In addition to the binding energies ($\Delta E_e$), the zero-point energies ($\Delta E_0$), enthalpies ($\Delta H$), and Gibbs free energies ($\Delta G$) at 298 K and 1 atm are also reported in this paper for both B3LYP and MP2 theoretical methods.

Average relativistic effective potential (AREP) and spin-orbit (SO) operators for the chemical elements of the second transition row have been published in Ref. [87], in which particular attention was focused on the portioning of the core and valence space, and Gaussian basis sets with contraction coefficients for the lowest energy state of each atom were introduced. Discussions of the details and complete review were given in Ref. [88]. More general reviews on the subject of effective core potentials can be also found in the literature [89-90]. The effective core potential from Ref. [87] called CRENBL ECP was used in this report to calculate structural and energetic properties of (super)-molecules containing a heavy iodine atom possessing atomic number 53 in the periodic table of chemical elements by Dmitrii I. Mendeleev. All optimized (super)-molecular structures introduced in this report were drawn with the POSMOL [91] (POhang Science and technology MOLecular modeling software package).

## 3. Results and Discussion

Table 1 lists the binding energies of acidic super-molecules and figures 1 and 2 show the optimized structures of the super-molecules obtained at the B3LYP-DFT/aug-cc-*p*VDZ level of theory, except the acidic super-molecules containing a heavy iodine atom, for which the effective core potential called CRENBL ECP was used. Using the B3LYP-DFT method with the basis set of aug-cc-*p*VDZ quality, it was found that three possible isomers for the chemical formula $H_2SO_3$ can exist. The most stable of them called Sulfurous acid ($H_2SO_3$) possesses the lowest energy on the potential energy surface. Its molecular structure is shown in figure 1 with both hydrogen atoms rotated towards the third molecular oxygen, which is not coupled with a hydrogen atom. The second, less stable $H_2SO_3$ isomer has a ground state energy approximately 1 kcal/mol less and oppositely directed hydrogen atoms (see table 1 and figure 1). The Sulfonic acid, being the third possible $H_2SO_3$ isomer, is the most "unstable" with the local minimum energy value of by about – 625.0345324 Hartrees. This value is ~ 1 kcal/mol less. However, a Sulfonic acid single molecule can have the strongest coupling with a single water molecule ($H_2O$) with a 10% larger binding energy $E_B$ compared with the other two isomers. Indeed, it is possible to suggest that a 10% difference in energy can be treated as non-significant, and hence, one deals here with an isoenergetic case like that studied in Refs. [92, 93]. Using the Moeller-Plesset second-order perturbation theory (MP2), obtained binding energies are always larger than the corresponding energies calculated at the hybrid B3LYP-DFT level of theory. Moreover, some optimized structures shown in figure 1 cannot be found with the MP2-method, for instance, the structure shown in table 1 with the largest binding energy $E_B$ for the most stable $H_2SO_3$ isomer. This can be explained by the way that the potential energy surfaces (PESs) for both commonly-used methods can give even qualitatively different results concerning structure optimization. For Sulfonic acid, it was also found that the water oxygen can interact with the $H_2SO_3$ molecular hydrogen coupled with the sulfur atom. This is natural because an $H_2O$-molecule can readily interact with an $H_2S$-molecule [56] if a sulfur atom is placed instead of an oxygen atom in a water dimer. However, in the case of O=HS-interaction for the $H_2SO_3$-$H_2O$ complex system, twice the binding energy can be obtained compared with the relatively more simple system of O-HS-interaction: an $H_2S$-$H_2O$-system. It is noted that the following acids HF, HCl, HBr, and HI with simple chemical formulae containing no oxygen atoms were recently studied by Odde *et al.* in Ref. [94]. Also, table 2 lists vibrational frequencies for the Sulfurous acid super-molecular isomers at both the B3LYP and MP2 levels of theory. A guide on molecular spectroscopy, vibration and rotation spectra can be found in an excellent book on physical chemistry in Ref. [95].

**Table 1. The binding energies of acid-water super-molecules. The lowest binding energies are shown in bold. The MP2 energy calculations are also given for the corresponding B3LYP-DFT energies. Note that some MP2 structures for the B3LYP ones were not found, and hence, their energies were not introduced in this table. Here, the binding energies $\Delta E_e$, zero point energies $\Delta E_0$, thermal enthalpies $\Delta H$, and thermal free energies $\Delta G$ are given in kcal/mol. The CRENBL ECP effective core potential was used to calculate the energies of super-molecules containing a heavy iodine atom. In the first column of the table, the labels in the parentheses for the super-molecular isomers correspond to those for the super-molecular structures shown in figures 1 and 2**

| Name(*) | B3LYP | | | | MP2 | | | |
|---|---|---|---|---|---|---|---|---|
| | $\Delta E_e$ | $\Delta E_0$ | $\Delta H$ | $\Delta G$ | $\Delta E_e$ | $\Delta E_0$ | $\Delta H$ | $\Delta G$ |
| $H_2SO_4$-$H_2O$(1U11t) | **-11.13** | **-8.89** | **-9.56** | **-0.385** | -12.69 | **-10.40** | -11.05 | **-1.903** |
| $H_2SO_4$-$H_2O$(1U'11t) | -11.09 | -8.85 | -9.51 | -0.384 | -12.73 | -10.39 | **-11.06** | -1.879 |
| $H_2SO_4$-$H_2O$(1U11c) | **-10.04** | **-7.86** | **-8.49** | **0.37** | **-11.55** | **-9.34** | **-9.92** | **-1.15** |
| $H_2SO_4$-$H_2O$(1U12c) | -8.86 | -6.40 | -7.08 | 1.85 | -11.38 | -8.87 | -9.58 | -0.09 |
| $H_2SO_4$-$H_2O$(1U10c) | -1.56 | -0.69 | -0.24 | 3.74 | -2.74 | -1.73 | -1.40 | 3.92 |
| $H_2SO_3$-$H_2O$(1U12sp) | -9.36 | -6.71 | -7.55 | 1.95 | **-11.72** | **-8.93** | **-9.84** | **0.16** |
| $H_2SO_3$-$H_2O$(1U'11sp) | -9.23 | -6.87 | -7.60 | 1.71 | -10.64 | -8.26 | -8.95 | 0.23 |
| $H_2SO_3$-$H_2O$(1U01sp) | -2.48 | -1.24 | -1.08 | 4.49 | -3.73 | -2.47 | -2.36 | 3.82 |
| $H_2SO_3$-$H_2O$(1U00sp) | -0.09 | 0.48 | 1.14 | 4.92 | - | - | - | - |
| $H_2SO_3$-$H_2O$(1U11sp) | **-9.83** | **-7.34** | **-8.15** | **1.32** | - | - | - | - |
| $H_2SO_3$-$H_2O$(1U02sp) | -2.16 | -0.94 | -0.79 | 5.56 | -4.31 | -2.92 | -2.93 | 4.63 |
| $H_2SO_3$-$H_2O$(1U'01sp) | - | - | - | - | -3.73 | -2.46 | -2.35 | 3.73 |
| $H_2SO_3$-$H_2O$(1U02sa) | -6.84 | -4.65 | -5.14 | 3.68 | -8.96 | -6.67 | -7.22 | 1.85 |
| $H_2SO_3$-$H_2O$(1U'11sa) | -8.94 | -6.41 | -7.21 | 2.36 | -10.59 | -8.03 | -8.81 | 0.75 |
| $H_2SO_3$-$H_2O$(1U'02sa) | -6.82 | -4.61 | -5.10 | 3.71 | -8.91 | -6.59 | -7.15 | 1.89 |
| $H_2SO_3$-$H_2O$(1U11sa) | **-9.17** | **-6.58** | **-7.41** | **2.23** | **-10.92** | **-8.27** | **-9.09** | **0.56** |
| $H_2SO_3$-$H_2O$(1U11su) | **-11.08** | **-8.70** | **-9.46** | **-0.20** | **-12.50** | **-10.06** | **-10.80** | **-1.56** |
| $H_2SO_3$-$H_2O$(1U00su) | -5.26 | -4.25 | -3.97 | 2.03 | -6.06 | -5.07 | -4.73 | 0.77 |
| $H_3PO_4$-$H_2O$(1U11pp) | **-11.55** | **-8.97** | **-9.84** | -0.25 | **-14.09** | **-11.48** | **-12.30** | **-2.99** |
| $H_3PO_4$-$H_2O$(1U'11pp) | -11.29 | -8.84 | -9.59 | -0.32 | -13.89 | -11.32 | -12.08 | -2.65 |
| $H_3PO_4$-$H_2O$(1U"11pa) | -10.66 | -8.27 | -9.09 | 0.39 | -13.06 | -10.63 | -11.41 | -2.15 |
| $H_3PO_4$-$H_2O$(1U02pa) | -7.74 | -5.50 | -6.09 | 2.85 | -10.54 | -8.22 | -8.80 | 0.19 |
| $H_3PO_4$-$H_2O$(1U'11pa) | -10.88 | -8.38 | -9.22 | 0.46 | -13.34 | -10.80 | -11.59 | -2.07 |
| $H_3PO_4$-$H_2O$(1U11pa) | **-11.10** | **-8.59** | **-9.44** | **0.22** | **-13.58** | **-10.97** | **-11.80** | **-2.21** |
| $H_3PO_3$-$H_2O$(1U'11pt) | -7.56 | -5.10 | -5.76 | 3.46 | -12.32 | -9.69 | -10.50 | **-0.95** |
| $H_3PO_3$-$H_2O$(1U"01pt) | -3.32 | -1.75 | -1.68 | 4.33 | -6.99 | -5.27 | -5.35 | 0.98 |
| $H_3PO_3$-$H_2O$(1U'01pt) | -6.22 | -4.72 | -4.74 | **2.04** | - | - | - | - |
| $H_3PO_3$-$H_2O$(1U01pt) | -6.43 | -4.72 | -4.87 | 2.29 | -11.55 | -9.17 | -9.91 | -0.47 |
| $H_3PO_3$-$H_2O$(1U11pt) | **-7.58** | -5.08 | **-5.79** | 3.59 | **-12.45** | **-9.79** | **-10.62** | -0.93 |
| $H_3PO_3$-$H_2O$(1U02p) | -2.61 | -1.33 | -1.22 | 5.37 | -4.67 | -3.19 | -3.25 | 4.54 |

**Table 1. (Continued)**

| Name(*) | B3LYP | | | | MP2 | | | |
|---|---|---|---|---|---|---|---|---|
| | $\Delta E_e$ | $\Delta E_0$ | $\Delta H$ | $\Delta G$ | $\Delta E_e$ | $\Delta E_0$ | $\Delta H$ | $\Delta G$ |
| $H_3PO_3$-$H_2O$(1U01p) | -2.94 | -1.58 | -1.51 | 5.02 | -4.46 | -3.03 | -3.03 | 3.92 |
| $H_3PO_3$-$H_2O$(1U11p) | **-11.65** | **-8.95** | **-9.89** | **-0.03** | **-13.10** | **-10.41** | **-11.32** | **-1.59** |
| $H_3PO_3$-$H_2O$(1U'11p) | -11.47 | -8.82 | -9.73 | 0.08 | -12.93 | -10.26 | -11.15 | -1.36 |
| $H_3PO_3$-$H_2O$(1U11a) | **-6.71** | **-4.56** | **-5.14** | **3.61** | **-8.33** | -6.06 | **-6.68** | 2.43 |
| $H_3PO_3$-$H_2O$(1U'11a) | -6.66 | -4.54 | -5.08 | 3.79 | -8.32 | **-6.07** | **-6.68** | **2.41** |
| $H_3BO_3$-$H_2O$(1U11b) | -7.50 | -5.26 | -5.76 | 3.07 | -9.03 | -6.75 | -7.26 | 1.59 |
| $HBO_2$-$H_2O$(1U01b) | -9.77 | -7.72 | -8.14 | -1.09 | -10.21 | -8.09 | -8.53 | -1.43 |
| HCN-$H_2O$(1U'00n) | -3.64 | -2.24 | -2.44 | 3.43 | -4.41 | -3.01 | -3.20 | 2.64 |
| HCN-$H_2O$(1U00n) | **-5.09** | **-3.97** | **-3.98** | **1.97** | **-5.66** | **-4.50** | **-4.48** | **1.19** |
| $HNO_3$-$H_2O$(1U10n) | -1.96 | -0.82 | -0.57 | 4.91 | -2.94 | -1.78 | -1.52 | 3.05 |
| $HNO_3$-$H_2O$(1U11n) | **-9.57** | **-7.56** | **-8.01** | **0.59** | **-10.31** | **-8.15** | **-8.59** | **0.00** |
| $HNO_3$-$H_2O$(1U'11n) | -1.72 | -0.85 | -0.44 | 3.53 | - | - | - | - |
| $H_2S$-$H_2O$(s1U00) | **-2.62** | **-1.06** | **-1.22** | 4.22 | -3.20 | -1.90 | -1.84 | **3.41** |
| $H_2S$-$H_2O$(s1U'00) | -2.28 | -1.04 | -0.95 | **3.79** | **-3.43** | **-1.92** | **-2.02** | 3.56 |
| $H_2Se$-$H_2O$(se1U00) | **-2.39** | **-0.82** | **-0.99** | 4.59 | -2.88 | -1.70 | -1.54 | 3.47 |
| $H_2Se$-$H_2O$(se1U'00) | -1.74 | -0.72 | -0.47 | **3.73** | **-3.57** | **-2.11** | **-2.18** | **3.39** |
| HClO-$H_2O$(c1U00) | -2.43 | -1.26 | -1.13 | 5.33 | -3.24 | -2.11 | -1.96 | 4.53 |
| HClO-$H_2O$(c1U01) | **-7.18** | **-5.16** | **-5.52** | **1.36** | **-8.03** | **-5.92** | **-6.29** | **0.43** |
| HClO-$H_2O$(c1U'01) | -6.97 | -5.03 | -5.33 | 1.48 | -7.70 | -5.70 | -6.00 | 0.73 |
| HClO-$H_2O$(c1U"01) | -2.68 | -1.20 | -1.24 | 4.70 | - | - | - | - |
| $HClO_2$-$H_2O$(c'1U11) | **-9.87** | **-7.45** | **-8.21** | **1.17** | **-11.02** | **-8.49** | **-9.25** | **0.15** |
| $HClO_2$-$H_2O$(c'1U01) | -2.83 | -1.59 | -1.48 | 5.24 | -5.16 | -3.78 | -3.82 | 3.62 |
| $HClO_2$-$H_2O$(c'1U10) | -4.30 | -2.74 | -2.81 | 3.71 | -5.34 | -3.63 | -3.80 | 3.31 |
| $HClO_2$-$H_2O$(c"1U10) | -3.60 | -2.32 | -2.23 | 4.17 | -4.89 | -3.62 | -3.55 | 3.35 |
| $HClO_2$-$H_2O$(c"1U11) | -9.80 | -7.38 | -8.13 | 1.22 | -10.90 | -8.32 | -9.09 | 0.32 |
| $HClO_3$-$H_2O$(c1U20) | -3.64 | -2.56 | -2.33 | 4.09 | -5.43 | -4.40 | -4.18 | 2.24 |
| $HClO_3$-$H_2O$(c1U11) | **-9.38** | **-6.96** | **-7.74** | **1.60** | **-9.76** | **-7.17** | **-7.96** | **1.46** |
| $HClO_3$-$H_2O$(c'1U'01) | -3.34 | -2.08 | -1.98 | 5.11 | -6.60 | -5.07 | -5.25 | 2.71 |
| $HClO_3$-$H_2O$(c'1U'10) | -3.19 | -2.10 | -1.88 | 4.18 | - | - | - | - |
| $HClO_4$-$H_2O$(1U'c11) | -10.20 | -8.14 | -8.68 | -0.01 | -11.24 | -9.00 | -9.55 | -0.82 |
| $HClO_4$-$H_2O$(1Uc11) | **-10.61** | **-8.51** | **-9.02** | **-0.86** | **-11.79** | **-9.42** | **-10.02** | **-1.17** |
| $HClO_4$-$H_2O$(1Uc10) | -1.26 | -0.40 | 0.02 | 4.98 | - | - | - | - |
| HBrO-$H_2O$(b1U00) | -3.97 | -2.61 | -2.61 | 4.51 | -5.10 | -3.84 | -3.78 | 3.20 |
| HBrO-$H_2O$(b1U01) | **-6.70** | **-4.70** | **-5.02** | 1.80 | **-7.79** | **-5.77** | **-6.07** | **0.09** |
| HBrO-$H_2O$(b1U'01) | -6.56 | -4.66 | -4.92 | **1.77** | -7.55 | -5.58 | -5.85 | 0.79 |
| HBrO-$H_2O$(b1U"01) | -3.08 | -1.59 | -1.62 | 3.90 | -4.38 | -2.74 | -2.87 | 3.22 |
| $HBrO_2$-$H_2O$(b'1U11) | **-9.86** | **-7.48** | **-8.22** | **1.19** | **-11.62** | **-9.19** | **-9.92** | **-0.51** |
| $HBrO_2$-$H_2O$(b'1U01) | -3.75 | -2.44 | -2.40 | 4.87 | -6.42 | -4.99 | -5.08 | 2.81 |
| $HBrO_2$-$H_2O$(b'1U10) | -4.97 | -3.37 | -3.48 | 3.19 | -6.52 | -4.74 | -4.97 | 2.45 |
| $HBrO_2$-$H_2O$(b"1U10) | -5.21 | -3.61 | -3.78 | 4.15 | -6.87 | -5.40 | -5.49 | 2.24 |
| $HBrO_2$-$H_2O$(b"1U11) | -9.77 | -7.38 | -8.11 | 1.28 | -11.49 | -9.03 | -9.77 | -0.33 |

**Table 1. (Continued)**

| Name(*) | B3LYP | | | | MP2 | | | |
|---|---|---|---|---|---|---|---|---|
| | $\Delta E_e$ | $\Delta E_0$ | $\Delta H$ | $\Delta G$ | $\Delta E_e$ | $\Delta E_0$ | $\Delta H$ | $\Delta G$ |
| $HBrO_3$-$H_2O$(b1U20) | -5.88 | -4.35 | -4.44 | 3.39 | -7.96 | -6.50 | -6.58 | 1.23 |
| $HBrO_3$-$H_2O$(b1U11) | **-10.30** | **-7.77** | **-8.64** | **0.95** | **-11.47** | **-8.87** | **-9.71** | **-0.23** |
| $HBrO_3$-$H_2O$(b'1U'11) | -5.11 | -3.58 | -3.65 | 3.81 | -7.96 | -6.26 | -6.51 | 1.99 |
| $HBrO_4$-$H_2O$(1U'b11) | -11.08 | -8.79 | -9.57 | **-0.11** | -12.50 | -10.17 | -10.88 | -1.66 |
| $HBrO_4$-$H_2O$(1Ub11) | **-11.27** | **-8.90** | **-9.68** | -0.05 | **-12.99** | **-10.51** | **-11.26** | **-1.88** |
| HIO-$H_2O$(i1U00) | **-6.82** | **-5.27** | **-5.37** | 2.21 | **-7.29** | **-5.90** | -5.92 | 1.43 |
| HIO-$H_2O$(i1U01) | -5.96 | -3.99 | -4.27 | 2.53 | -7.14 | -5.20 | -5.44 | **1.03** |
| HIO-$H_2O$(i1U'01) | -5.86 | -4.03 | -4.23 | **2.20** | -7.04 | -5.19 | **-5.93** | 2.09 |
| HIO-$H_2O$(i1U"01) | -4.41 | -2.64 | -2.87 | 3.88 | -6.06 | -4.18 | -4.49 | 2.46 |
| $HIO_2$-$H_2O$(i'1U11) | **-10.00** | **-7.70** | **-8.37** | **0.92** | **-13.04** | **-10.62** | **-11.37** | **-1.84** |
| $HIO_2$-$H_2O$(i'1U01) | -5.80 | -4.17 | -4.36 | 3.79 | -8.53 | -6.88 | -7.13 | 1.29 |
| $HIO_2$-$H_2O$(i'1U10) | -6.85 | -5.12 | -5.33 | 2.00 | -8.97 | -7.07 | -7.40 | 0.45 |
| $HIO_2$-$H_2O$(i"1U10) | -8.39 | -6.40 | -6.87 | 2.07 | -10.09 | -8.22 | -8.62 | 0.15 |
| $HIO_2$-$H_2O$(i"1U11) | -9.93 | -7.64 | -8.29 | 0.96 | -12.88 | -10.45 | -11.20 | -1.66 |
| $HIO_3$-$H_2O$(i1U20) | -9.35 | -7.40 | -7.76 | 0.83 | -10.91 | -8.99 | -9.36 | -0.90 |
| $HIO_3$-$H_2O$(i1U11) | **-11.37** | **-8.62** | **-9.67** | **0.79** | **-13.08** | **-10.28** | **-11.31** | **-1.04** |
| $HIO_3$-$H_2O$(i'1U'11) | -7.74 | -5.52 | -6.08 | 3.55 | -10.25 | -8.05 | -8.64 | 1.04 |
| $HIO_4$-$H_2O$(1U'i11) | -11.894 | -9.507 | -10.349 | -0.696 | **-13.449** | -11.013 | **-11.850** | -2.663 |
| $HIO_4$-$H_2O$(1Ui11) | **-11.895** | **-9.513** | **-10.351** | **-0.717** | **-13.449** | **-11.017** | **-11.850** | **-2.733** |

(*) – in the parentheses ($NUN_1N_2$"$L$") of the isomer names, $N$ is for the number of water molecules in the super-molecules, $U$ means "undissociated" molecules in the super-molecules, $N_1$ is for the number of molecular oxygen atoms interacting with a water molecule, and $N_2$ is for the number of molecular oxygen-hydrogen group interacting with a water molecule; "$L$" is for identification of acid individuality in a super-molecule, for instance, "$L$" = "$t$" for the trans $H_2SO_4$, "$L$" = "$c$" for the cis $H_2SO_4$, etc. Note that for the Cl-, Br-, and I-containing super-molecules, "$L$" = "$c$", "$b$", and "$i$" can be before $N$ and after $U$. This is also true for $H_2S$-$H_2O$ and $H_2Se$-$H_2O$.

Previous DFT studies to obtain the binding energy of acid-water super-molecules were carried out for the "trans" Sulfuric acid $H_2SO_4$-$H_2O$(1U11t) as well as for the "cis" Sulfuric acid isomers, $H_2SO_4$-$H_2O$(1U11c) and $H_2SO_4$-$H_2O$(1U12c). The binding energies from Ref. [50] (namely, – 12.80 kcal/mol and even – 15.10 kcal/mol for the "trans" $H_2SO_4$-$H_2O$(1U11t) isomer, and – 10.50 kcal/mol and – 12.40 kcal/mol for the "cis" $H_2SO_4$-$H_2O$(1U12c) isomer) are significantly higher than those listed in table 1. However, the works [47, 49] give relatively the same binding energy value of – 9.70 kcal/mol compared with that listed in table 1 for the "cis" $H_2SO_4$-$H_2O$(1U11c) isomer. According to the results in Ref. [56] for the Sulfur-containing acid with the simplest chemical formula $H_2S$, both the B3LYP-DFT and MP2 studies give relatively the same binding energy values of – 1.39 kcal/mol and – 1.34 kcal/mol, respectively, which completely disagree with the data for the $H_2S$-$H_2O$(s1U00) isomer listed in table 1. In addition, the theoretical results of this report indicate that for the two possible $H_2S$-$H_2O$ super-molecular isomers, the DFT calculations demonstrate the most stable optimized structure for the $H_2S$-$H_2O$(s1U00) isomer, while the MP2 calculations illuminate the most stable optimized structure for the second $H_2S$-$H_2O$(s1U'00) isomer. The same was obtained for the $H_2Se$-$H_2O$ isomers to support it. Concerning the HCN-acid study in previous works, the obtained binding energy values calculated in 2006 by Malaspina *et al.* [71] are positive for the HCN-$H_2O$(1U'00n) and HCN-$H_2O$(1U00n) isomers compared with the corresponding negative values listed in table 1 for the HCN-$H_2O$ isomers. Note that the table 1 negative values for the HCN-$H_2O$ isomers are situated significantly below the corresponding large positive values recently calculated by Malaspina *et al.* [71].

Also, some theoretical investigations of the extremely strong acid $HBF_4$ (probably, the second strongest known acid which cannot be isolated: soluble in water and diethyl ether) were carried out in this work. This fluorinated acid is very useful because it can be soluble in organic solutions in addition to aqueous ones. One can find the recent theoretical and experimental studies of fluorinated acids in Ref. [96]. The B3LYP-DFT study of this theoretical work obtained the optimized structure for an $HBF_4$ single molecule with a very long F–B bond length ~ 2.52 Å between the two stable molecules HF and $BF_3$, of which the latter has a short F–B bond length ~ 1.33 Å. Hence, this agrees with the fact that an $HBF_4$ single molecule is unstable, and therefore, the binding energy for the acid-water interaction probably cannot be evaluated. Adding a single water molecule to this "unstable" $HBF_4$ system resulted in significant shortening of the very long F–B bond length down to ~ 2.10 Å. This can be understood as an $HBF_4$-$H_2O$ system becoming more stable. It was also found that an additional $H_2O$-molecule for the $HBF_4$-$H_2O$ system can dissociate $HBF_4$, and additional two $H_2O$ molecules can give a significantly longer F–B bond length larger than 2.52 Å for the HF-$BF_3$ ($HBF_4$) complex.

**Table 2. The frequencies [$cm^{-1}$] for the super-molecular isomers of water-coupled Sulfurous acid $H_2SO_3$, using the B3LYP-DFT/aug-cc-*p*VDZ level of theory (the corresponding MP2/aug-cc-*p*VDZ frequencies are shown in the parentheses). The labels of optimized structures listed in table 1 for the $H_2SO_3$–$H_2O$ super-molecular isomers are given in the parentheses of the title row. These frequencies were recalculated from the original harmonic frequencies applying the scale factors such as 0.970 for the B3LYP-DFT and 0.959 for the MP2 taken from http://srdata.nist.gov/cccbdb/vibscalejust.asp.**

| $H_2SO_3$–$H_2O$(1U'11sp) | $H_2SO_3$–$H_2O$(1U02sp) | $H_2SO_3$–$H_2O$(1U12sp) | $H_2SO_3$–$H_2O$(1U01sp) | $H_2SO_3$–$H_2O$(1U00sp) | $H_2SO_3$–$H_2O$(1U11sp) |
|---|---|---|---|---|---|
| 48 (41) | 28.6 (59) | 36 (71) | 16 (12) | 17 | 46 |
| 158 (153) | 29.0 (65) | 150 (165) | 20 (45) | 26 | 167 |
| 188 (175) | 89 (111) | 190 (188) | 95 (94) | 38 | 191 |
| 333 (326) | 269 (277) | 387 (365) | 309 (304) | 149 | 360 |
| 397 (394) | 324 (333) | 416 (411) | 350 (320) | 305 | 403 |
| 462 (460) | 380 (383) | 460 (461) | 393 (390) | 368 | 463 |
| 771 (767) | 692 (686) | 716 (718) | 706 (698) | 705 | 770 |
| 1015 (1030) | 1042 (1050) | 1045 (1072) | 1043 (1051) | 1045 | 1019 |
| 1099 (1108) | 1065 (1064) | 1122 (1119) | 1062 (1055) | 1059 | 1102 |
| 3530 (3519) | 3589 (3527) | 3475 (3433) | 3589 (3526) | 3587 | 3513 |
| 3596 (3539) | 3671 (3632) | 3518 (3498) | 3637 (3616) | 3674 | 3590 |
| 3752 (3730) | 3768 (3750) | 3755 (3730) | 3766 (3759) | 3779 | 3744 |

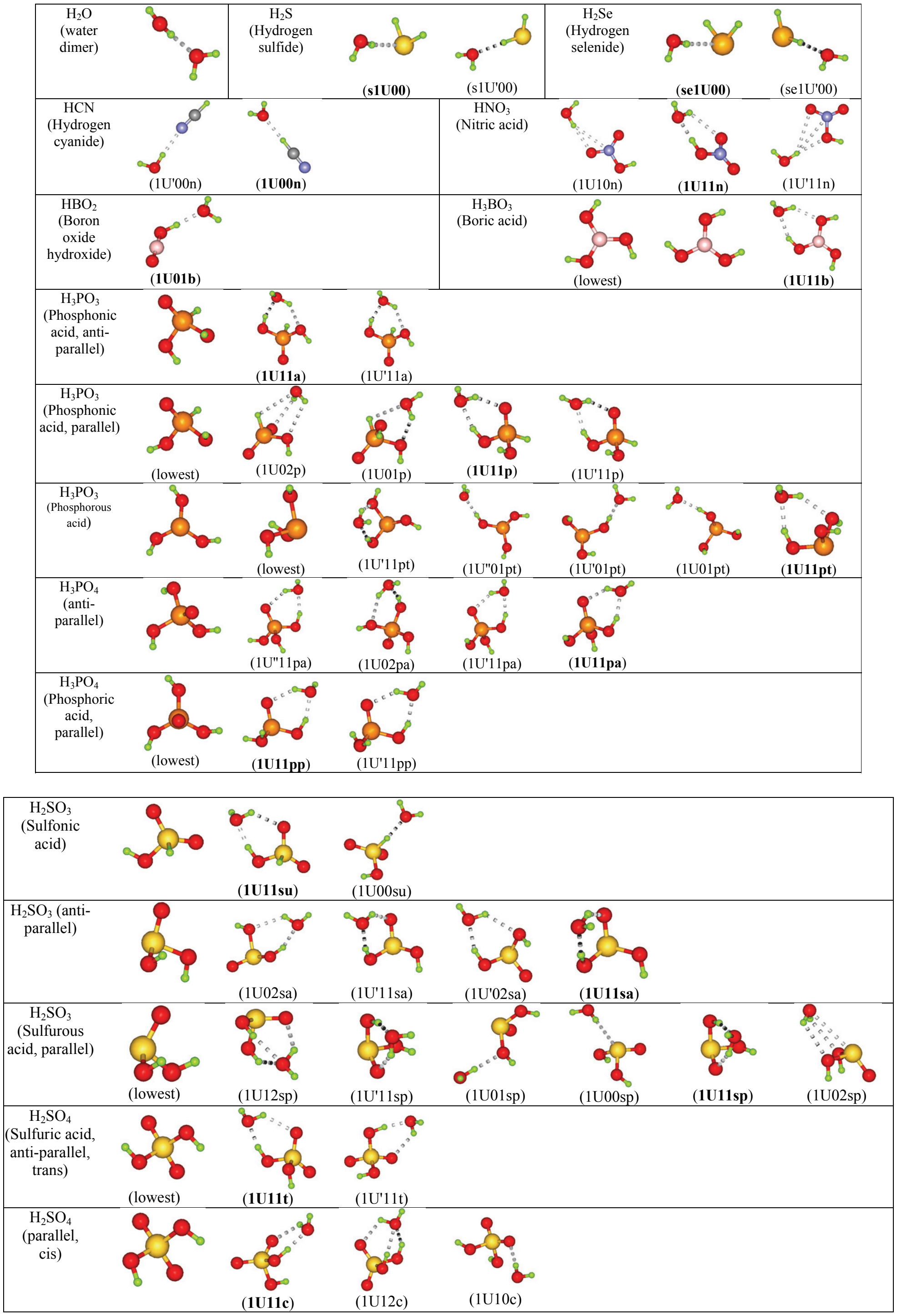

**Figure 1.** The system configurations for optimized geometries (B3LYP/aug-cc-*p*VDZ). The label under each super-molecular structure corresponds to that in the first column of table 1

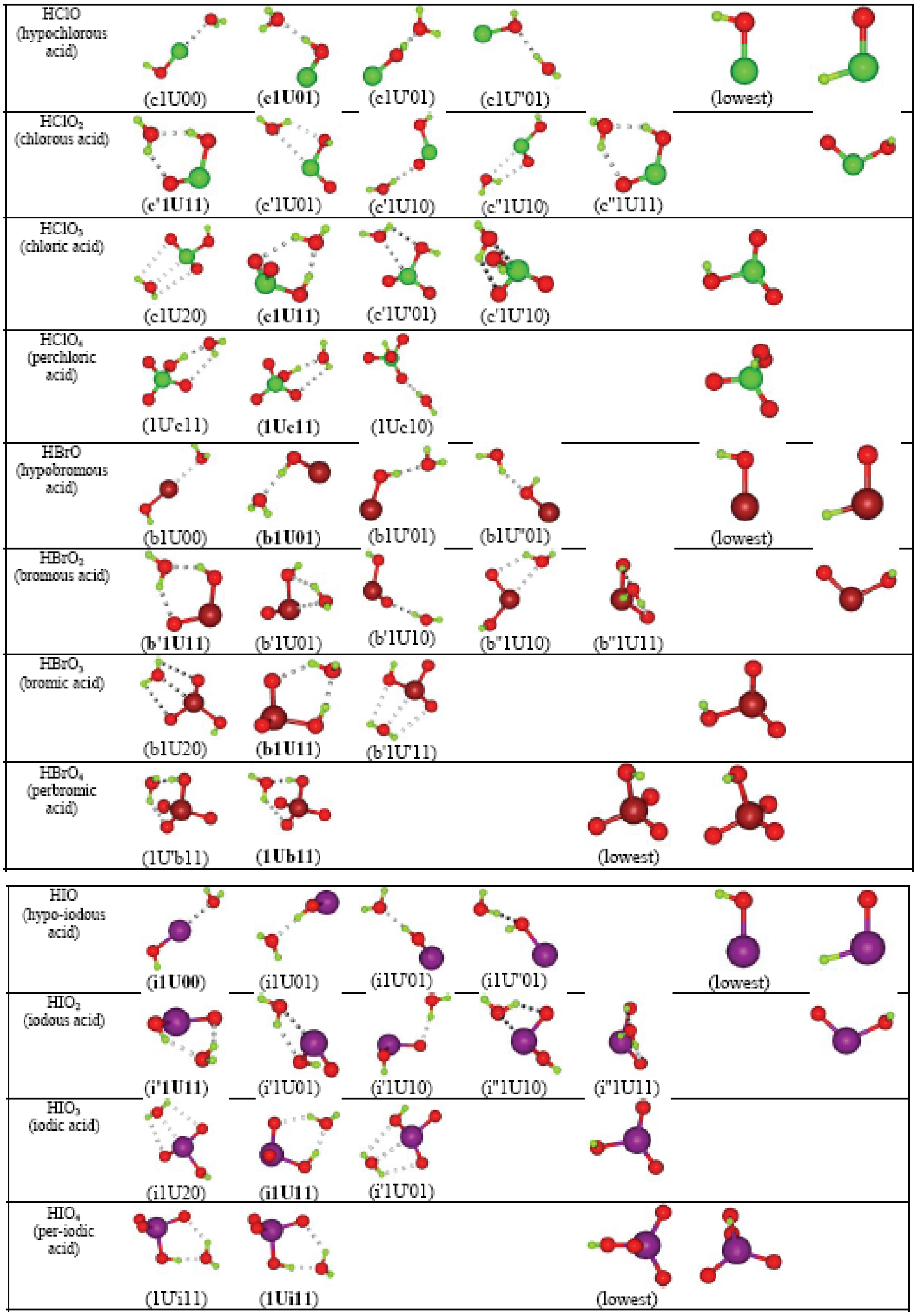

**Figure 2.** The system configurations for optimized geometries (B3LYP/aug-cc-*p*VDZ). The CRENBL ECP effective core potential was used for a heavy iodine atom. The label under each super-molecular structure corresponds to that in the first column of table 1

It is possible to continue the discussions about the calculations done with the B3LYP-DFT and MP2 methods using the basis set of aug-cc-*p*VDZ quality. Note that the energy of a water monomer is $E_w$(B3LYP-DFT/aug-cc-*p*VDZ) = – 76.4446427 atomic units (Hartrees) and the energy for the MP2 calculations is $E_w$(MP2/aug-cc-*p*VDZ) = – 76.2609098 Hartrees. It is noted that the energy difference between the parallel-like and anti-parallel-like hydrogen orientations for $H_3PO_4$ is as high as the following value: $\Delta E_M$($H_3PO_4$, B3LYP-DFT/aug-cc-*p*VDZ) = $E_M$(anti-parallel) – $E_M$(parallel) ~ 0.90 kcal/mol. The same value for Phosphonic acid $H_3PO_3$ is as follows: $\Delta E_M$($H_3PO_3$, B3LYP-DFT/aug-cc-*p*VDZ) = $E_M$(anti-parallel) – $E_M$(parallel) = ~ 1.79 kcal/mol. This difference between the parallel-like and anti-parallel-like hydrogen orientations for two possible hydrogen orientations of Sulfuric acid ($H_2SO_4$) is as high as the following value: $\Delta E_M$($H_2SO_4$, B3LYP-DFT/aug-cc-*p*VDZ) = $E_M$(anti-parallel) – $E_M$(parallel) ~ – 1.13 kcal/mol. The same value for Sulfurous acid $H_2SO_3$ is as follows: $\Delta E_M$($H_2SO_3$, B3LYP-DFT/aug-cc-*p*VDZ) = $E_M$(anti-parallel) – $E_M$(parallel) ~ 0.93 kcal/mol. Using the MP2 calculations, these values for $H_3PO_4$ are as follows: $\Delta E_M$($H_3PO_4$, MP2/aug-cc-*p*VDZ) ~ 0.98 kcal/mol; the value for Phosphonic acid $H_3PO_3$ is as follows: $\Delta E_M$($H_3PO_3$, MP2/aug-cc-*p*VDZ) ~ 1.88 kcal/mol; for Sulfuric acid $H_2SO_4$, one can calculate $\Delta E_M$($H_2SO_4$, MP2/aug-cc-*p*VDZ) ~ – 1.17 kcal/mol; the Sulfurous acid $H_2SO_3$ value is as follows: $\Delta E_M$($H_2SO_3$, MP2/aug-cc-*p*VDZ) ~ 0.81 kcal/mol.

The interaction energy $E_B$ of different molecules with a single water molecule is written in formula (1) below, where $E_T$ represents the total energy consisting of the energy of a sample molecule $E_M$ plus the energy of a single water molecule $E_w$. $E_C$ is the corrected energy with the basis set superposition error (BSSE) correction giving the energy error $E_{BSSE}$. The interaction energy $E_{BC}$ can be also written as follows:

$$E_B = E_T - E_M - E_w \text{ and } E_{BC} = E_C - E_M - E_w \qquad (1)$$

The binding energy, using the values of the $E_{BSSE}$, can be calculated with the following formula:

$$E_{B\Delta} = (E_B + E_{BC} \pm E_{BSSE})/2 \qquad (2)$$

It is noted that in order to calculate the binding energies $E_B = \Delta E_e$ listed in table 1, the energies $E_M$ of the most stable molecular isomers were used where it was applicable. Using the energies $E_M$ of the less stable molecular isomers can result in larger values of the binding energies in equations (1) and (2). For example, using the $H_2SO_4$-cis energy value of – 700.2787452 Hartrees instead of the $H_2SO_4$-trans lowest energy value of – 700.2805479 Hartrees, it can be obtained that $\Delta E_e(E_M = -700.2787452) \sim -11.17$ kcal/mol for the $H_2SO_4$-$H_2O$(1U11c) isomer can be situated slightly below $\Delta E_e(E_M = -700.2805479) \sim -11.13$ kcal/mol for the $H_2SO_4$-$H_2O$(1U11t) isomer. The same situation can occur for the $H_2SO_3$-$H_2O$ super-molecular isomers: $\Delta E_e(E_M = -625.0776524) \sim -9.83$ kcal/mol for the $H_2SO_3$-$H_2O$(1U11sp) isomer, $\Delta E_e(E_M = -625.0761673) \sim -10.11$ kcal/mol for the $H_2SO_3$-$H_2O$(1U11sa) isomer (see the corresponding table value for comparison) and $\Delta E_e(E_M = -625.0345324) \sim -11.08$ kcal/mol for the $H_2SO_3$-$H_2O$(1U11su) isomer.

Using formula (2) for the binding energy with the BSSE correction, the maximum $E_{BSSE}$ was calculated for the "cis" $H_2SO_4$-$H_2O$(1U12c) isomer with $E_{B\Delta} \sim -(9.55 \pm 0.40)$ kcal/mol at the B3LYP-DFT/aug-cc-*p*VDZ level of theory. For the calculations at the MP2/aug-cc-

*p*VDZ level of theory, the maximum BSSE correction was obtained for the same super-molecular isomer: $E_{B\Delta}$ ~ – (11.42 ± 1.13) kcal/mol. That is very close to the energy value of $E_{B\Delta}$ ~ – (12.22 ± 1.12) kcal/mol for the $H_3PO_4$-$H_2O$(1U'11pa) isomer. On the other hand, the minimum BSSE correction was obtained with the energy value of $E_{B\Delta}$ ~ – (1.63 ± 0.10) kcal/mol for the $H_2Se$-$H_2O$(se1U'00) isomer at the B3LYP-DFT/aug-cc-*p*VDZ level and $E_{B\Delta}$ ~ – (2.85 ± 0.35) kcal/mol for the $H_2S$-$H_2O$ (s1U00) isomer at the MP2/aug-cc-*p*VDZ level. It is very interesting that the $H_2SO_3$-$H_2O$(1U00sp) super-molecular isomer possesses the smallest binding energy with the BSSE correction energy value approximately 1.5 times larger than the binding energy value: $E_{B\Delta}$ ~ – (– 0.10 ± 0.15) kcal/mol at the B3LYP-DFT/aug-cc-*p*VDZ level. Applying higher level of theory with B3LYP-DFT/aug-cc-*p*VQZ [83, 84, 97], the following value was calculated for the $H_2SO_3$-$H_2O$(1U00sp) super-molecular isomer: $E_{B\Delta}$ ~ – (0.05 ± 0.02) kcal/mol. This is already 2.5 times smaller than the calculated binding energy.

In the calculations of the binding energy with the B3LYP method of density-functional theories for the $H_2SO_3$-$H_2O$(1U00sp) super-molecular isomer, the energies $E_M(H_2SO_3)$ = – 625.2020613 Hartrees and $E_w$ = – 76.47224 Hartrees were first obtained for the augmented basis set of aug-cc-*p*VQZ (Quadrupole Zeta) quality. Note that there is a dramatic situation for comparison of the B3LYP-DFT and MP2 theories studying the $H_2SO_3$-$H_2O$ isomers, because the corresponding MP2 isomers cannot be found for the $H_2SO_3$-$H_2O$(1U00sp) isomer with the lowest binding energy and the $H_2SO_3$-$H_2O$(1U11sp) isomer with the largest binding energy. This can be explained by a gross difference in the potential energy surfaces of two methods. Also, the corresponding MP2 isomers cannot be found for the B3LYP-DFT $HNO_3$-$H_2O$(1U'11n) super-molecular isomer and for some others, see table 1 and the figures. However, the reverse situation can also occur when the MP2 $H_2SO_3$-$H_2O$(1U'01sp) isomer listed in table 1 can exist with no corresponding optimized structure at the B3LYP-DFT/aug-cc-*p*VDZ level of theory.

For the studied acids containing chlorine, bromine, and iodine atoms (see table 1 and figure 2) it was obtained that the largest BSSE correction energy values among the Cl-containing acids are $E_{BSSE}/2$ ~ 0.25 kcal/mol for the $HClO_2$-$H_2O$(c'1U11), $HClO_4$-$H_2O$(1U'c11), and $HClO_4$-$H_2O$(1Uc11) isomers at the B3LYP-DFT/aug-cc-*p*VDZ level, as well as $E_{BSSE}/2$ ~ 1.0 kcal/mol for the $HClO_4$-$H_2O$(1Uc11) isomer at the MP2/aug-cc-*p*VDZ level. Among the Br-containing acids, the maximum values of $E_{BSSE}/2$ were calculated for the $HBrO_2$-$H_2O$(b''1U10) and $HBrO_4$-$H_2O$(1Ub11) isomers with $0.5E_{BSSE}$(B3LYP-DFT/aug-cc-*p*VDZ) ~ 0.28 kcal/mol and for the $HBrO_3$-$H_2O$(b'1U'11) isomer with $0.5E_{BSSE}$(MP2/aug-cc-*p*VDZ) ~ 1.23 kcal/mol.

For the studied acids containing a heavy iodine atom, for which the CRENBL ECP effective core potential was chosen in the DFT and MP2 calculations, the maximum ($E_{BSSE}/2$)-values are as follows: $0.5E_{BSSE}$(B3LYP-DFT/aug-cc-*p*VDZ) ~ 0.61 kcal/mol for the HIO-$H_2O$(i1U00) and $HIO_3$-$H_2O$(i'1U'11) super-molecular isomers and $0.5E_{BSSE}$(MP2/aug-cc-*p*VDZ) ~ 1.68 kcal/mol for the $HIO_2$-$H_2O$(i'1U11) isomers. It is obvious that the DFT-maximum ($E_{BSSE}/2$)-values are relatively the same for the Cl- and Br-containing acids which are two times smaller than the values for the I-containing acids, using the ECP for a heavy iodine atom in the calculations. However, this difference is not dramatic for the MP2 calculations: the MP2-maximum ($E_{BSSE}/2$)-value of the Br-containing acids is ~ 23% larger than that for the Cl-containing acids, but the maximum value of the I-containing acids is only ~ 36% larger than that for the Br-containing acids.

## 4. Conclusions

In this report, the structural and energetic characteristics of the interaction of single acid molecules with a single water molecule have been investigated. Both commonly-used theoretical methods, the Becke's three-parameter exchange potential with the Lee-Yang-Parr correlation functional and the second-order Møller-Plesset perturbation theory, have shown a good agreement for finding optimized structures and calculations of binding energies, using the same Dunning's basis set of aug-cc-*p*VDZ quality and the CRENBL ECP effective core potential for molecules containing a heavy iodine atom. However, some MP2 optimized structures of the acid-water super-molecules cannot be found for the corresponding B3LYP-DFT calculations. That can be explained by the uniqueness of each potential energy surface for the theoretical methods. That is even dramatic for the Sulfurous acid-water ($H_2SO_3$-$H_2O$) super-molecular isomers, because the corresponding MP2 optimized structures cannot be found for the B3LYP-DFT ones with the smallest and largest binding energies. The corrected energies with the basis set superposition error (BSSE) correction were also calculated for the studied acid super-molecules.

## Acknowledgements

It is a pleasure to acknowledge the Pohang University of Science and Technology for support from a POSTECH grant. Also, the author is grateful to Professor K.S. Kim, S. Karthikeyan, and M. Kolaski for fruitful discussions and their solid interest in the work.